\def\year{2017}\relax
\documentclass[letterpaper]{article}
\usepackage{aaai17}
\usepackage{times}
\usepackage{helvet}
\usepackage{courier}

\usepackage{color}
\usepackage{graphicx}
\usepackage{amsmath}
\usepackage{amssymb}
\usepackage{caption}
\usepackage{subcaption}
\usepackage{multirow}
\usepackage{threeparttable}

\makeatletter
\newcommand*{\rom}[1]{\expandafter\@slowromancap\romannumeral #1@}
\makeatother

\begin{document}
\title{A multi-task learning model for malware classification with useful file access pattern from API call sequence}
\author{%xxxxxxx\\
Xin Wang \and Siu Ming Yiu\\
Department of Computer Science , The University of Hong Kong, Hong Kong\\
%\{xxxx\}@cs.hku.hk
\{xwang, smyiu\}@cs.hku.hk
}
\maketitle
\begin{abstract}
Based on API call sequences, semantic-aware and machine learning (ML) based malware classifiers can be built for malware detection or classification. Previous works concentrate on crafting and extracting various features from malware binaries, disassembled binaries or API calls via static or dynamic analysis and resorting to ML to build classifiers. However, they tend to involve too much feature engineering and fail to provide interpretability. We solve these two problems with the recent advances in deep learning: 1) RNN-based autoencoders (RNN-AEs) can automatically learn low-dimensional representation of a malware from its raw API call sequence. 2) Multiple decoders can be trained under different supervisions to give more information, other than the class or family label of a malware. Inspired by the works of document classification and automatic sentence summarization, each API call sequence can be regarded as a sentence. In this paper, we make the first attempt to build a multi-task malware learning model based on API call sequences. The model consists of two decoders, one for malware classification and one for $\emph{file access pattern}$ (FAP) generation given the API call sequence of a malware.  We base our model on the general seq2seq framework. Experiments show that our model can give competitive classification results as well as insightful FAP information. 
\end{abstract}

\section{Introduction}
Malware continues to be one the the big security threats for both the Internet and computing devices. It can be used for espionage, advertisements promotion, ransom demand and other unauthorised activities on your networks and systems. Due to the ubiquity of malware, automatic tools are usually deployed for malware detection or classification. So many ML algorithms have been applied to the classification problems of malwares.

Classification problems of malwares can be divided into two types: (\rom{1}) malware detection, which is a binary classification problem and decides whether a sample is benign or malicious; (\rom{2}) malware classification, which is a multi-class classification problem and outputs the family label of a sample known to be malicious. We refer to them as type \rom{1} and type \rom{2} respectively.

There are generally two kind of approaches to analyze malwares that are used to build ML-based malware classifiers: static analysis and dynamic analysis. Static analysis     examines the binary executables directly or after disassembling without really executing them. Diverse static features can be used to build malware classifiers, such as PE header information, n-grams at different granularity levels \cite{masud2008scalable}, global descriptors of malware image \cite{nataraj2011malware}, section entropy, etc. Usually, compositions of various static features are used to get high accuracy \cite{ahmadi2016novel}, which consequently lead to high-dimensional binary sparse feature vectors. Random projections and PCA \cite{dahl2013large} or feature selection \cite{lin2015feature} are often performed to reduce the dimensions of the input space. However, the problem is that extractions of some static features are already very time consuming and memory intensive. Even worse, various obfuscation techniques \cite{you2010malware} make static analysis difficult nowadays and evasive techniques are available to defeat it \cite{sikorski2012practical}. Dynamic analysis observes the behaviors of a malware usually by actually executing it in a sandbox environment. Malwares are then correlated based on the similarity of their behaviors. Two typical methods of dynamic analysis are control flow analysis and API call analysis. API call information can be extracted from both static analysis and dynamic analysis.  Earlier works tend to use a simple frequency representation \cite{tian2010differentiating} \cite{shankarapani2010kernel} of the API calls, whose drawback is evident that  API calls in one sequence are treated individually and isolately. The sequentiality of the API calls is such a very important feature that should be considered. Some works extract API call semantics \cite{zhang2014semantics} or control flow information \cite{christodorescu2005semantics} into graph representations. However, it involves too much manual tweak on graph matching and sophiscated feature engineering. Feature engineering can be hard because it requires specific domain knowledge to design helpful features and involves burdensome deployment and testing. Given the huge amount and ever increasing diversity of the malwares, it is important to build scalable model that can learn features of malware automatically. In this paper, we propose a model that learns representations of malware samples in an unsupervised way.

 %Many sophisticated works \cite{zhang2014semantics} \cite{kolbitsch2009effective}  in dynamic or static analysis for malware classification have been proposed, while the problem is that they involve too much feature engineering. 
 
Furthermore, another big problem of previous works on malware detection or classification is the lack of interpretability. For malware detection, a type \rom{1} classifier marks a suspicious sample as benign or malicious without give an understandable reason of the marking. As for malware classification, a type \rom{2} classifier outputs a family label.  However, sometimes just knowing the family label is not very helpful or even meaningless. For example, packing is often used by malwares to hide their real payload, which theoretically can be used by any malwares. Though a classifier can tell you that a malware is packed, it tells nothing about the payload of the packed malware. The payload matters because it is the payload that cause substantial harms to a system and a typical classifier is not able to tell you what the payload does. Due to the flexibility of the payload of a malware, our focus should come to the payload. 

\iffalse For example, a rootkit \cite{wiki:Rootkit} is a malware that can have access to computer or areas of its software that would not otherwise be allowed  while hiding its existence or the existence of other software at the same time. When a malware is identified to be a rootkit, what it does to your system is still unknown.\fi Even worse, when it comes to zero-day \cite{wiki:Zero-day} malwares, a type \rom{2} classifier will make a mistake because it can only output predefined and already known family labels. So we want to ask the question of can we provide more interpretability for classification problems of malware? As an answer to this question,  we propose to generate an FAP, which is a brief description of file access related behaviors, for each sample rather than just giving the class or family label. 

We propose a more interpretable model, multi-task malware learning model, that can be used for malware classification and FAP generation at the same time. Apart from classification, we define an FAP generation task which is combined with classification to provide interpretable information. We construct our model based on the multi-task seq2seq model \cite{luong2015multi}. Classification and FAP generation, though quite different, are both defined as seq2seq problems in our model.

In summary, this paper makes the following contributions:
\begin{itemize}
\item We propose a novel multi-task malware learning model on raw malware API call sequences. First, low dimensional representations of malware API call sequences are learned by RNN-AE in an unsupervised manner. Then multiple decoders, malware classifier and FAP generator, can be trained under the corresponding supervisions. To the best of our knowledge, this is the first time that multi-task learning is applied to malware learning to provide more interpretability.
\item Apart from a malware classifier, we propose a decoder for FAP generation. Inspired by the works of automatic sentence summarization \cite{rush2015neural}\cite{nallapatiabstractive} in Natural Language Processing (NLP), we formulate the FAP generation problem, which can be seen as  automatic summarization of API call sequences. As far as we know, this is also the first attempt.
\item In our model, we reformulate the malware classification problem with RNN as a special case of seq2seq problem whose output length equals to 1. Instead of obtaining the feature vectors of malware by training an RNN to predict the next API call \cite{pascanu2015malware}, we think that learning representations with RNN-AE is more natural and intuitive because RNN-AE has been widely used for representation learning in many other tasks.
\end{itemize}

\section{Preliminaries}

\subsubsection{Recurrent Neural Nerworks}
Unlike traditional neural network that assume all inputs are independent of each other, a recurrent neural network (RNN) is able to deal with sequential data with variable length and has shown its power in many NLP tasks.  The idea is an RNN maintains a hidden state which can be seen as the memory of the network. At each time step $t$, the hidden state $h_t$ is updated by:$$h_t=f(h_{t-1}, x_t),$$ where $f$ is an activiation function that usually provides nonlinearity, and $x_t$ is the input at time step $t$. At each time step, an RNN performs the same calculations on different inputs with the same shared parameters. 

An RNN can effectively learn the conditional distribution $p(\mathbf{Y}|\mathbf{X})$ where output sequence $\mathbf{Y} = (y_1, ... , y_{T^{'}})$ and input sequence $\mathbf{X} = (x_1, ... , x_{T})$. The length of input $T$ and that of output $T^{'}$ can be different. Combined with hidden state update equation above, The conditional probability distribution $p(\mathbf{Y}|\mathbf{X})$ can be unrolled to: $$p(y_1, ... , y_{T^{'}}| x_1, ... , x_{T}) =  \prod_{t=1}^{T^{'}} p(y_t | h_{t-1}, y_1, ... , y_{t-1})$$ 

Each $p(y_t | h_{t-1}, y_1, ... , y_{t-1})$ is a softmax distribution over all the input symbols.

\subsubsection{Autoencoders}
Autoencoder (AE) is a special neural network which tries to reconstruct its input. An AE consists of an encoder and a decoder, where the encoder map the input to a low-dimensional fixed-length vector from which the decoder recover the input as output. The most important feature of AE is that it learns in an unsupervised way.
 %\textcolor{red}{*To be added*}

%\subsubsection{Automatic summarization}
%Automatic summarization takes as input a text document and create a summary that retains the most important ideas of the original document. Automatic summarization if part of machine learning and data mining and has a long history in research community. In this paper, we specifically mean the recent works based on seq2seq model, which are natural extensions of neural machine translation.

\section{Multi-task Malware Learning Model}

The basic seq2seq model can be transformed into multi-task seq2seq learning model \cite{luong2015multi}. The multi-task extension can improve the performance of seq2seq model on machine translation centered tasks. According to the number of encoders or decoders, three settings are available in MTL seq2seq model: one-to-many, many-to-one, many-to-many. Our model employs a one-to-many setting, consisting of one encoder for representation learning and two decoders for classification and FAP generation respectively (see Fig. 1).

\begin{figure}[!htbp]
    \centering
    \includegraphics[width=0.48\textwidth, height=4cm]{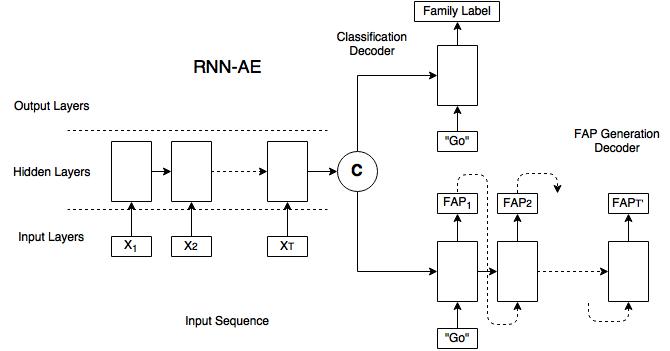}  
    \caption{Multi-task Malware Learning Model}
\end{figure}

%\subsubsection{Learning based on API call sequences}

%API call analysis is one of the major dynamic analysis methods and often used to build malware classifiers. . % the selection of critical APIs based on assumptions of API frequency and evasive code can be embeded to affect the graph matching.  

\subsubsection{Representation learning on API call sequence}
API\footnote{API in this paper means Windows kernel API by default.} call sequences are usually collected by hooking \cite{wiki:Hooking} while running the samples in a sandbox environment. Compared to other methods, one advantage of classifying malwares using API call sequences is that they are inherently semantic-aware since every single API call is an exact action performed by malware, e.g. creation, read, write, modification and deletion of files or registry keys. One thing we should emphasize is that the semantic information of an API call sequence does not only lies in each single API call, but also lies in the sequence itself.  

RNNs are known for their ability to capture the long term features in sequential or time-series data. An RNN trained to predict next API call can be applied for malware classification  \cite{pascanu2015malware}. Hidden states of the model are used as feature vectors to train a separate classifier. While in our model, we reformulate the malware classification problem with RNNs  as a special sequence to sequence learning \cite{sutskever2014sequence} (also known as seq2seq) problem whose output length equals to one. In a seq2seq model, a sequence is read by an RNN encoder into fixed-length hidden state, which then can be fed to an RNN decoder to predict the output sequence. Both the length of input and output can be variable, making the seq2seq a very general and powerful model for many applications, e.g. machine translation \cite{sutskever2014sequence}, text summerization \cite{nallapatiabstractive}, sentiment analysis \cite{dai2015semi}.
   
RNN-AE maps an input API call sequence to a fixed-length vector $C$, which is usually a low-dimensional representation of an input API call sequence. %Then another RNN initialize its internal state with $C$ and decode it to the target variable-length sequence (see Fig. 1).

%\textcolor{red}{*To be added*}
\subsubsection{Multiple Decoders}
The classification task in our model trains a type \rom{2} classifier and performs malware classification on samples known to be malicious, while the FAP generation task trains an API call sequence summarizer and outputs file access summaries of samples. A decoder shares the same structure as the RNN-AE used for representation learning, and its hidden state is initialized with the accumulated internal state $C$ (see Fig. 1).  A decoder starts with feeding a start symbol \textquotedblleft GO\textquotedblright, and generates a distribution. The symbol whose index corresponds to the highest probability in the distribution is the output symbol. Then the symbol generated is fed to the decoder iteratively until the target sequence is fully generated. 

Decoders are trained to minimize the cross-entropy over the sequences. Given two discrete probability distribution $p$ and $q$, the cross entropy between them is 
\begin{equation}
H(p, q) = -\sum_{x} p(x) \log q(x)
\end{equation}
. In the language model, cross entropy measures how accurate the model is in predicting the test data. $p$ is the true distribution, where the entry corresponding to the true target equals to 1 and others equals to 0. $q$ is the learned distribution. The closer $p$ and $q$ are, the smaller the cross-entropy is. The loss function is the full cross-entropy over the test dataset 
\begin{equation}
L(\theta) = -\frac{1}{N}\sum^{N}_{i=1} \sum^{L}_{j=1} \log q_{\theta}( y_{ij}| x_i, C) 
\end{equation}
 where N and L are the number of the test dataset and the length of the target sequence respectively, and $\theta$ corresponds to the model parameters.

\section{Evaluations}
We evaluate the accuracy of our model on a public malware API call sequence dataset \cite{web:dataset} at different granularity levels. We select two datasets which include 7430 samples for coarse-grained evaluation and  4932 samples for fine-grained evaluation (see Tab. 1 and Tab. 6). Malwares whose families are unknown or whose corresponding families do not have enough samples for training are dropped. They are split randomly into train, validation and test datasets, containing 75\%, 5\%, 20\% samples respectively. %The distributions of all the four families in train, validation, test sets are considered uniform.

 %All the samples are from 4 families: adware \cite{wiki:Adware}, packed \cite{wiki:Packed}, worm \cite{wiki:Worm} and trojan-fakeav \cite{web:Trojan-fakeav}.

\begin{table}
\begin{center}
\begin{tabular}{| l | c | c | c |}
\hline
Family & Number & Ratio & Mean of Length\\
\hline
Trojan-fakeav & 3247 & 43.7\% & 228\\
\hline
Adware & 2354 & 31.7\% & 203\\
\hline
Packed & 964 & 13.0\% & 320\\
\hline
Worm & 865 & 11.6\% & 224\\
\hline
Total & 7430 & 100.0\% & 235\\
\hline
\end{tabular}
\caption{Dataset Summarization}
\end{center}
\end{table}

\subsection{Preprocessing}
Supervisions are necessary for the training of decoders. For classification, class labels are already available in the dataset. As to the FAP generation, we first give our definition of FAP and then briefly describe how we extract an FAP for each malware.

\textbf{Definition of FAP} : Assume $S = \{s_1, s_2, ..., s_n\}$ is a set of file access related APIs, which is called FAP set, and $l \in \mathbb{Z}$, then we say $p = |s|^l$ is an FAP of length $l$, where $s\in S$.

We select seven file access related APIs from Windows kernel APIs as our FAP set. Different Windows kernel APIs that perform the same function are mapped to one API (see Table 2). For example, both $\textbf{CopyFile}$ and $\textbf{CopyFileEx}$ perform the function that copy a file, and the difference lies in whether the file already exists. Each function usually have two names for Unicode with $\textbf{W}$ as suffix and for ANSI with $\textbf{A}$ as suffix respectively.  All of them are mapped to $\textbf{CopyFile}$ in our FAP set. %We also add a "blank" symbol to the FAP set for padding.

We employ a simple way to extract FAPs. We first set the length of our FAP to be $l_p = |S|$, where $|S|$ denotes the number of elements in set $S$ \iffalse and the the "blank" symbol does not count\fi. Then we generate a binary representation for each malware, \iffalse, which is very similar to the bag of words (BOW) representation. The binary representation of a malware \fi  which is a binary vector $v$ with length $l_p$. For the $i$-th malware, \iffalse its binary representation stores in binary vector $v_i$ and\fi $v_i[j] = 1$ if the $j$-th API in FAP set can be extracted from the $i$-th malware API call sequence, otherwise $v_i[j] = 0$, where $j = 1,...,l_p$. We call concatenation of the elements in $S_i$ the FAP of the $i$-th malware, where $S_i$ is subset of $S$ corresponding to the binary vector $v_i$. \iffalse For convenience in experiments, we pad the FAP of a malware with $"[blank]"$ symbol until its length equals to $l_p$.\fi For example, if FAP set $S = \{a, b, c, d\}$ and $v_i = [1,0,1,1]$, then we get $S_i = \{a,c,d\}$ and $p_i = "acd"$ after concatenation. %Then the final FAP will be $"acd[blank]"$.

\begin{table}

\resizebox{0.45\textwidth}{!}{
	\centering
	\begin{minipage}{0.5\textwidth}
	\begin{tabular}{ l | c}
  	\hline			
  	APIs in FAP set & Original APIs\\
  	\hline
  	CreateFile & CreateFileA,CreateFileW \\
  	ReadFile & ReadFile \\
  	GetTempFileName & GetTempFileNameA,GetTempFileNameW \\
  	SetFileAttributes & SetFileAttributesA,SetFileAttributesW\\
  	WriteFile & WriteFile \\
  	CopyFile & CopyFileA,CopyFileExW \\
  	DeleteFile & DeleteFileA,DeleteFileW\\
  	\hline
	\end{tabular}
	\caption{Mapping of FAP set and original APIs}
	\end{minipage}
}
\end{table}

\subsection{Model setup}

As our model is a multi-task model based on basic seq2seq model, we first experiment on seq2seq model for classification and FAP generation as baselines. Then we verify our multi-task malware learning model on classification and FAP generation tasks respectively. There are many variants RNN units since the advent of LSTM \cite{hochreiter1997long}. \iffalse Instead of using the standard LSTM unit, \fi We use GRU as our default RNN unit, which has been shown to be more computationally efficient than standard LSTM without identifiable loss of performance \cite{chung2014empirical}.  The standard experiment (AE + Decoder) trains on the train dataset and decodes on the test dataset. We can also train on the full dataset (AE(full) + Decoder) because representation learning is unsupervised. We also evaluate on the bidirectional extension of the our GRU-based model with train dataset (bAE + Decoder) and full dataset (bAE(full) + Decoder). For the standard unidirectional RNNs, the output at each time step depends on the inputs at that time step and before. The bidirectional extension allow an RNN has access to both the inputs in the past and future. %We test our model with different maximum sequence length, and the number of layers of the RNN is adjusted accordingly. 
%\textcolor{red}{*To be adjusted*}
%We set the batch size to 32 and batches are also shuffled randomly before feeding. We optimize with momentum with.  The default number of epochs for training of RNN-AE is 50 and for training of decoders, the number of epochs is set to 20.

\subsection{Coarse-grained Evaluations and Results}
\begin{figure}[!ht]
    \centering
        \includegraphics [width=0.75\linewidth] {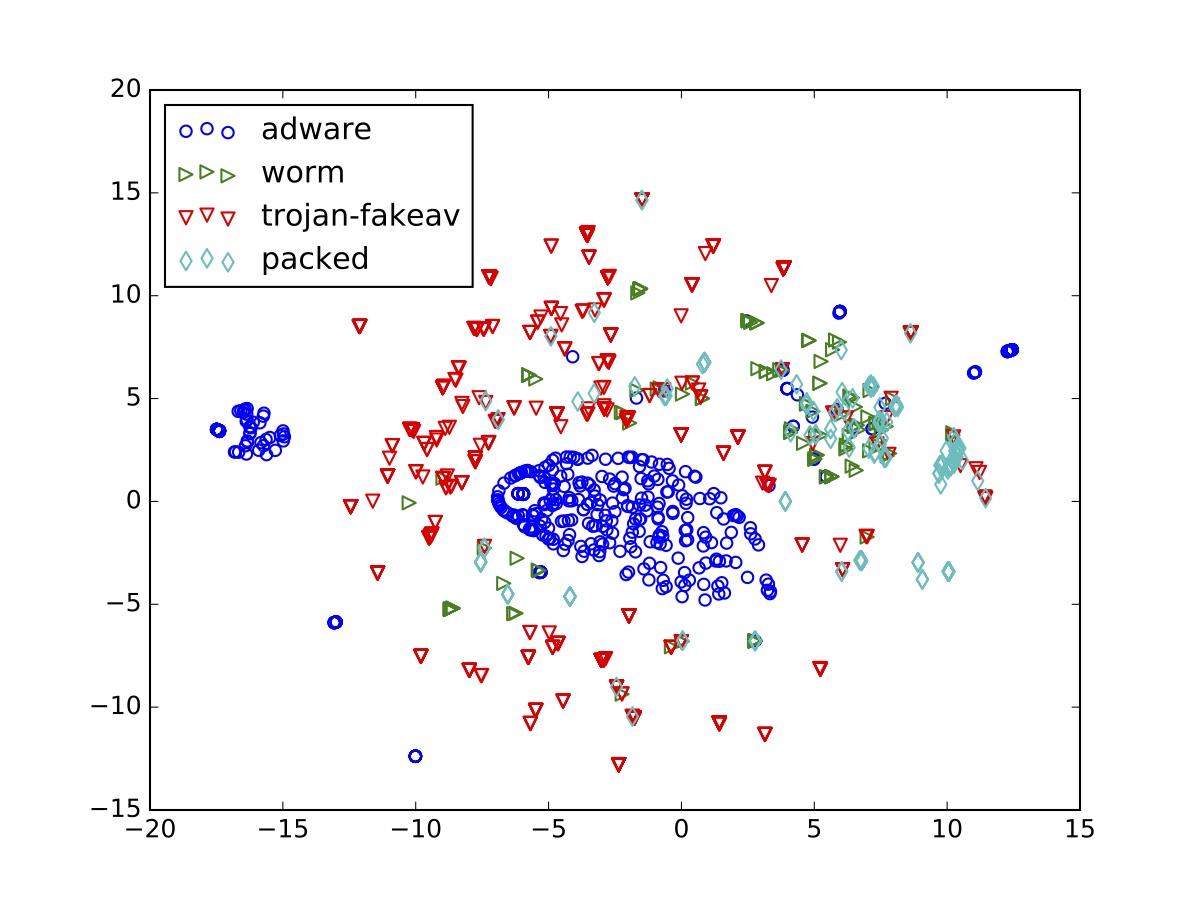}
    \caption{Visualization of Malware Features in Coarse-Grained Evaluation}
    \label{fig: coarse-grained tsne}
\end{figure}

\begin{table}
\centering
\resizebox{0.45\textwidth}{!}{\begin{minipage}{0.5\textwidth}
\begin{tabular}{ l | c | c | c| c}
  \hline			
  \multirow{2}{3cm}{Model} & \multicolumn{2}{c|}{$F_3$} & \multicolumn{2}{c}{$F_3$ + \textbf{packed}} \\
  \cline{2-5}
  	&Train & Test & Train & Test\\ 
  \hline
  seq2seq classification & 99.6\% & 98.2\% & 99.6\% & 96.5\% \\
  \hline
  AE+Decoder & 99.6\% & \textbf{97.9\%} & 99.5\% & 96.5\%  \\
  AE(full)+Decoder & 99.5\% & 97.7\% & 99.4\% & \textbf{96.6\%}  \\
  \hline
  bAE+ bDecoder & 99.6\% & 97.5\% & 99.6\% & 96.3\% \\
  bAE(full)+ bDecoder & 99.5\% & 97.6\% & 99.5\% & 96.3\% \\
  \hline
\end{tabular}
\caption{Performance on classification task}
\end{minipage}}
\end{table}

%\subsubsection{File Access Pattern Generation}
We experiment on the ground truth. All FAP candidates are mapped to indices (see Tab. 5). We first evaluate on samples from \textbf{trojan-fakeav}, \textbf{adware}, \textbf{worm}, which we denote with $F_3$, and then on samples from $F_3$ and \textbf{packed}. The reason behind this is that we find classification performance on samples from $F_3$ and packed decreases evidently compared to the performance on samples from $F_3$. We visualize the feature vectors in tsne \cite{maaten2008visualizing}, a method to visualising high-dimensional vectors (see Fig. 2). Adware samples form two clear clusters, while samples from other families are very scattered and some of them are highly interweaved. A feature vector, learned from the API call sequence, can be seen as a $\emph{behavioral aggregation}$ of a malware. In Fig. 2, we can see samples from different families may share very similar behavioral aggregation, while samples from the same family may share quite different behavioral aggregation. That is what a classifier alone can not tell. However, an FAP generator is able to extract the pattern that a malware access the file system regardless of which family the malware belongs to. Unlike classification, the performance of FAP generation does not fluctuate on different datasets. The malwares from different families exhibit obviously different dominant FAPs (see Fig. 3).  In our empirical experiments, the trainings with full data or bidirectional extension of GRU do not bring evident increase in performance. As to the connections between some specific FAPs and the malware families, we refer to the following section \textbf{Case Studies} for the detailed analyses. 

\begin{table}
\centering
\resizebox{0.45\textwidth}{!}{\begin{minipage}{0.5\textwidth}
\begin{tabular}{ l | c | c| c |c }
  \hline			
  \multirow{2}{3cm}{Model} & \multicolumn{2}{c|}{$F_3$} & \multicolumn{2}{c}{$F_3$ + \textbf{packed}} \\
  \cline{2-5}
  	&Train & Test & Train & Test\\ 
  \hline
  seq2seq summarization & 99.7\%& 98.7\% &99.5\% & 98.6\% \\
  \hline
  AE+Decoder & 98.8\% & \textbf{98.4\%} & 98.8\% & \textbf{98.3\%}  \\
  AE(full)+Decoder & 98.7\% & \textbf{98.4\%} & 98.6\% & 98.2\%  \\
  \hline
  bAE+ bDecoder & 98.8\% & 98.3\% & 98.6\% & 98.2\% \\
  bAE(full)+ bDecoder & 98.7\% & 98.2\% & 98.5\% & 98.1\% \\
  \hline
\end{tabular}
\caption{Performance on FAP generation task}
\end{minipage}}
\end{table}

\subsection{Case Studies}
We analyze the possible correlations between the FAP of a malware and its family in this section. Presumably, malwares from different families may exhibit different FAPs, and those who are in the same family is supposed to have similar FAPs to some extent.  %Though the FAP of a malware does not necessarily determine the family the malware belongs to, it can tell what the malware does to your file system.  %
\begin{figure*}[!ht]
    \centering
    \begin{subfigure}[b]{0.24\textwidth}
        \includegraphics[width=\textwidth]{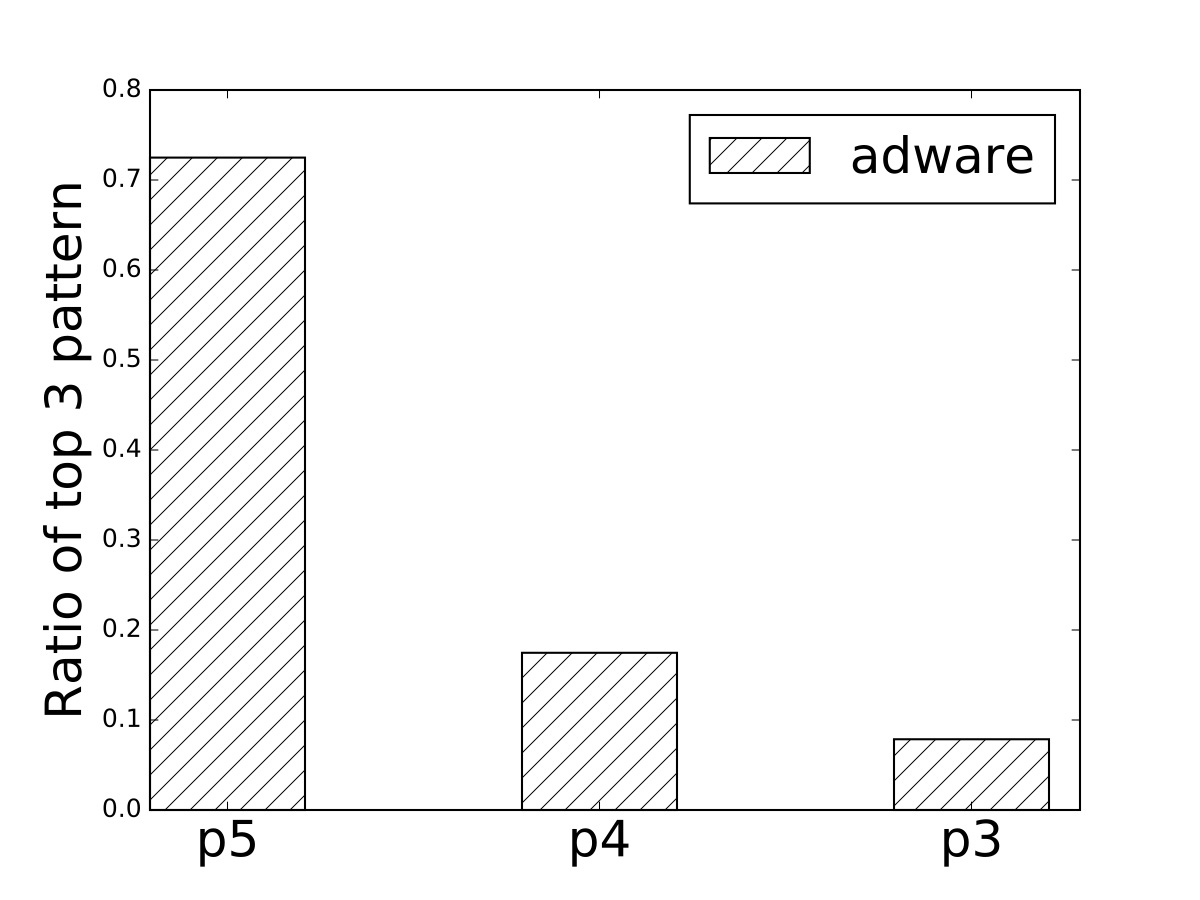}
        \caption{adware}
        \label{fig:adware}
    \end{subfigure}
    \begin{subfigure}[b]{0.24\textwidth}
        \includegraphics[width=\textwidth]{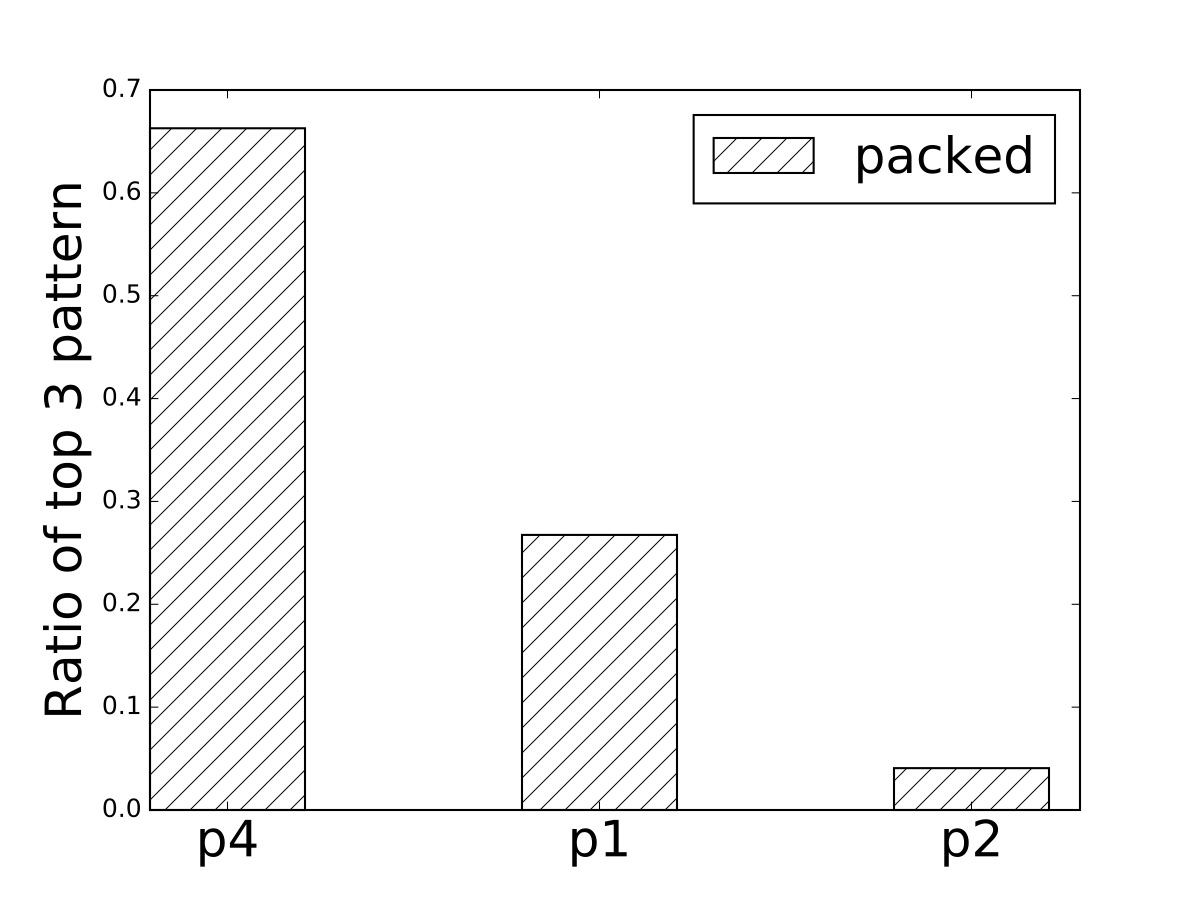}
        \caption{packed}
        \label{fig:packed}
    \end{subfigure}
    \begin{subfigure}[b]{0.24\textwidth}
        \includegraphics[width=\textwidth]{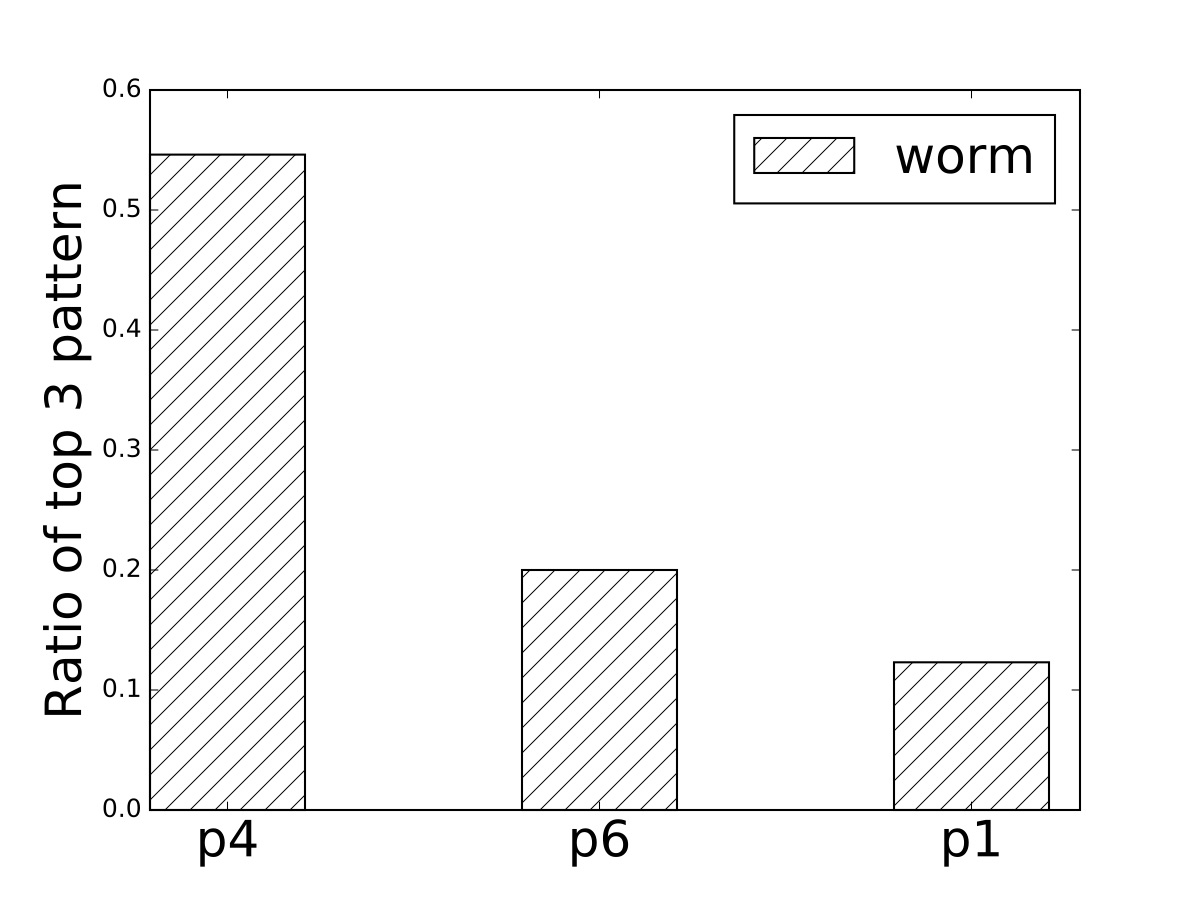}
        \caption{worm}
        \label{fig:worm}
    \end{subfigure}
    \begin{subfigure}[b]{0.24\textwidth}
        \includegraphics[width=\textwidth]{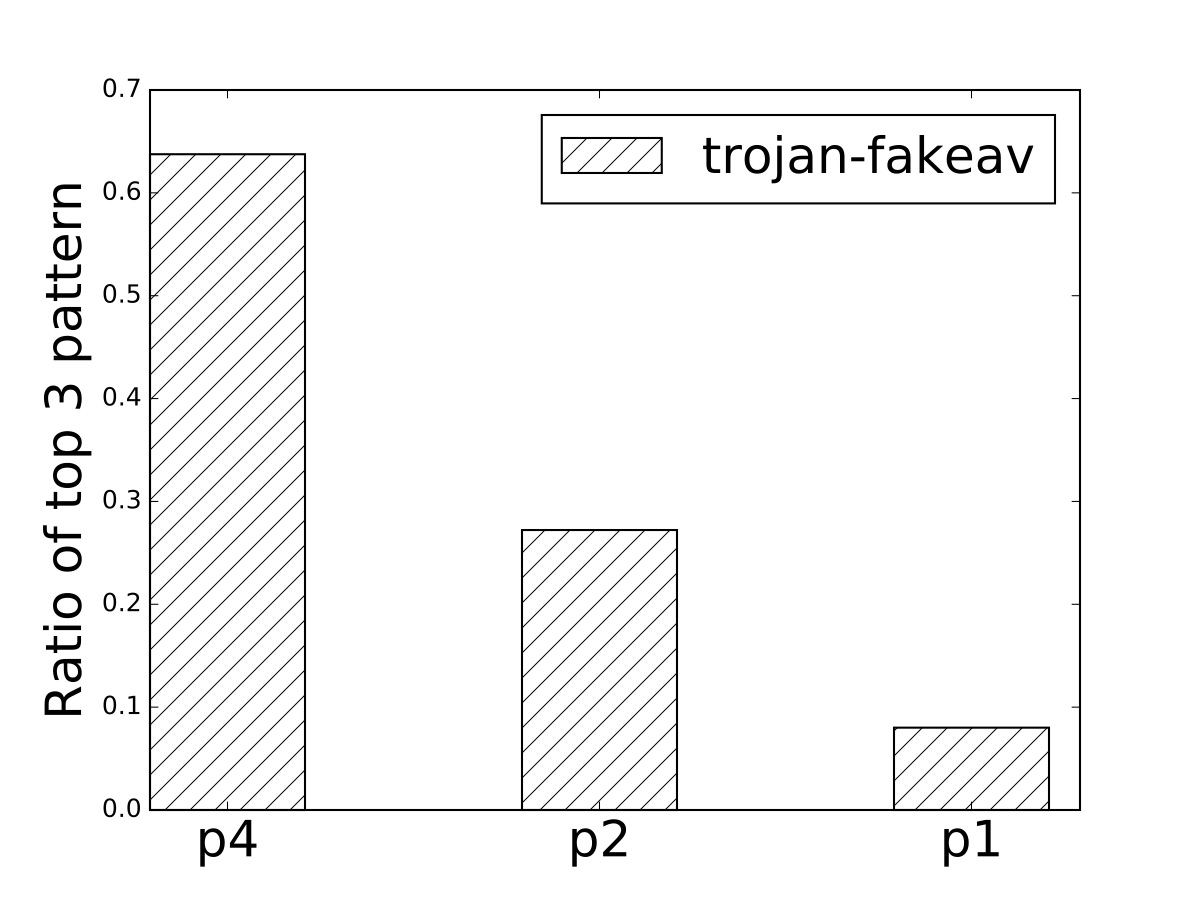}
        \caption{trojan-fakeav}
        \label{fig:trojan-fakeav}
    \end{subfigure}
    \caption{Top 3 FAPs for each family}
    \label{fig: summarization}
\end{figure*}
\begin{table}[!ht]
\begin{center}
\resizebox{0.35\textwidth}{!}{\begin{minipage}{0.4\textwidth}
\begin{tabular}[width=0.3\linewidth]{ l | c }
  \hline			
  FAP & ID \\
  \hline
  CreateFile\_WriteFile & p1 \\
  CreateFile\_ReadFile & p2 \\
  CreateFile\_WriteFile\_ReadFile & p3 \\
  CreateFile & p4 \\
  CreateFile\_ReadFile\_GetTempFileName\_\\SetFileAttributes\_DeleteFile\_WriteFile & p5 \\
  CreateFile\_WriteFile\_CopyFile & p6\\
  \hline
\end{tabular}
\end{minipage}}
\caption{FAP mapping list}
\end{center}
\end{table}

\begin{description}
\item [Worm] A worm is a kind of malware that usually spreads by replicating itself via network. One of the typical actions a worm will perform is "Copy" \cite{analysis:Dabber-C} \cite{analysis:Doomjuice-B}. We can find that the ratio of FAP p6, including "CopyFile",  is considerable in the worm samples, yet is ignorable in samples from other family (see Fig. 3(c) and Tab. 5).

\begin{figure}[!ht]
    \centering
    \begin{subfigure}[b]{0.23\textwidth}
        \includegraphics[width=\textwidth]{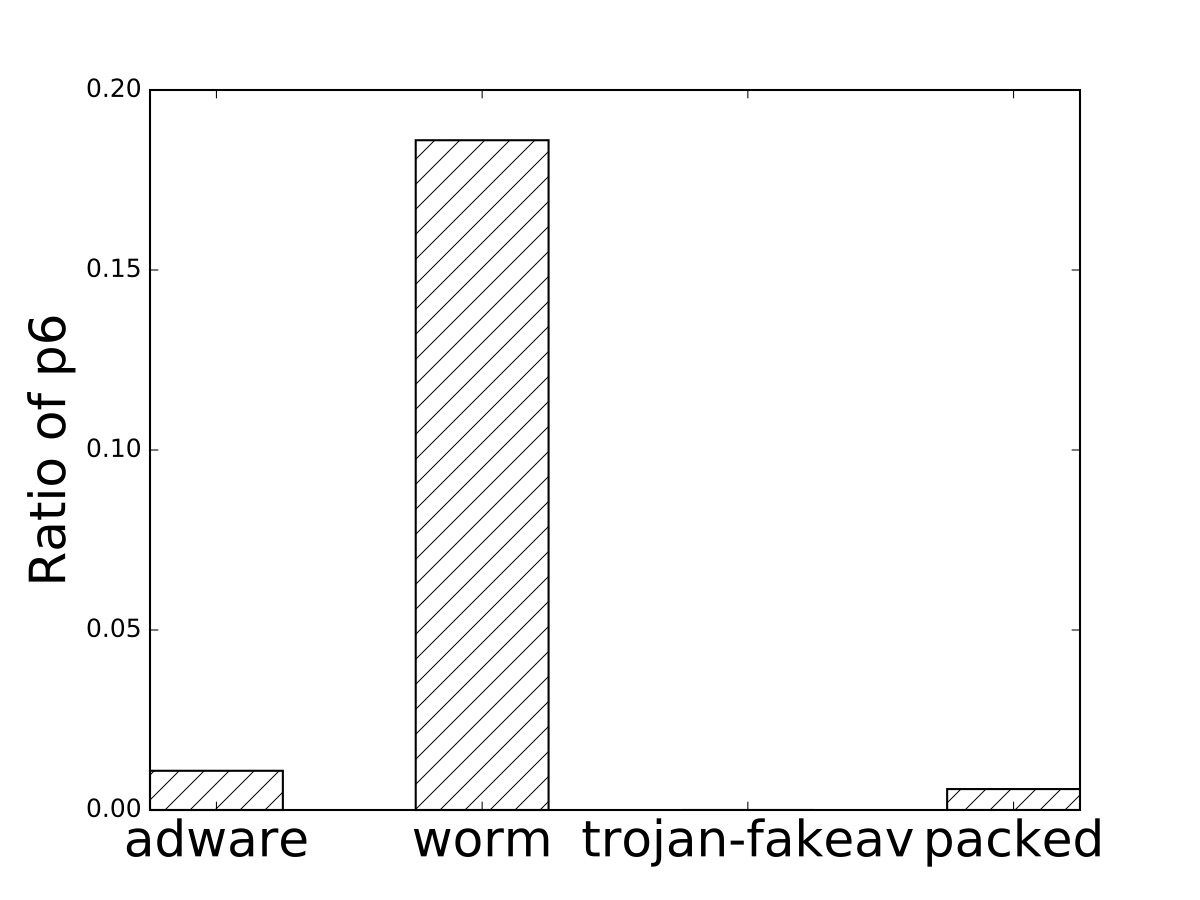}
        \caption{Distribution of FAP p6}
        \label{fig:worm}
    \end{subfigure}
    \begin{subfigure}[b]{0.23\textwidth}
        \includegraphics[width=\textwidth]{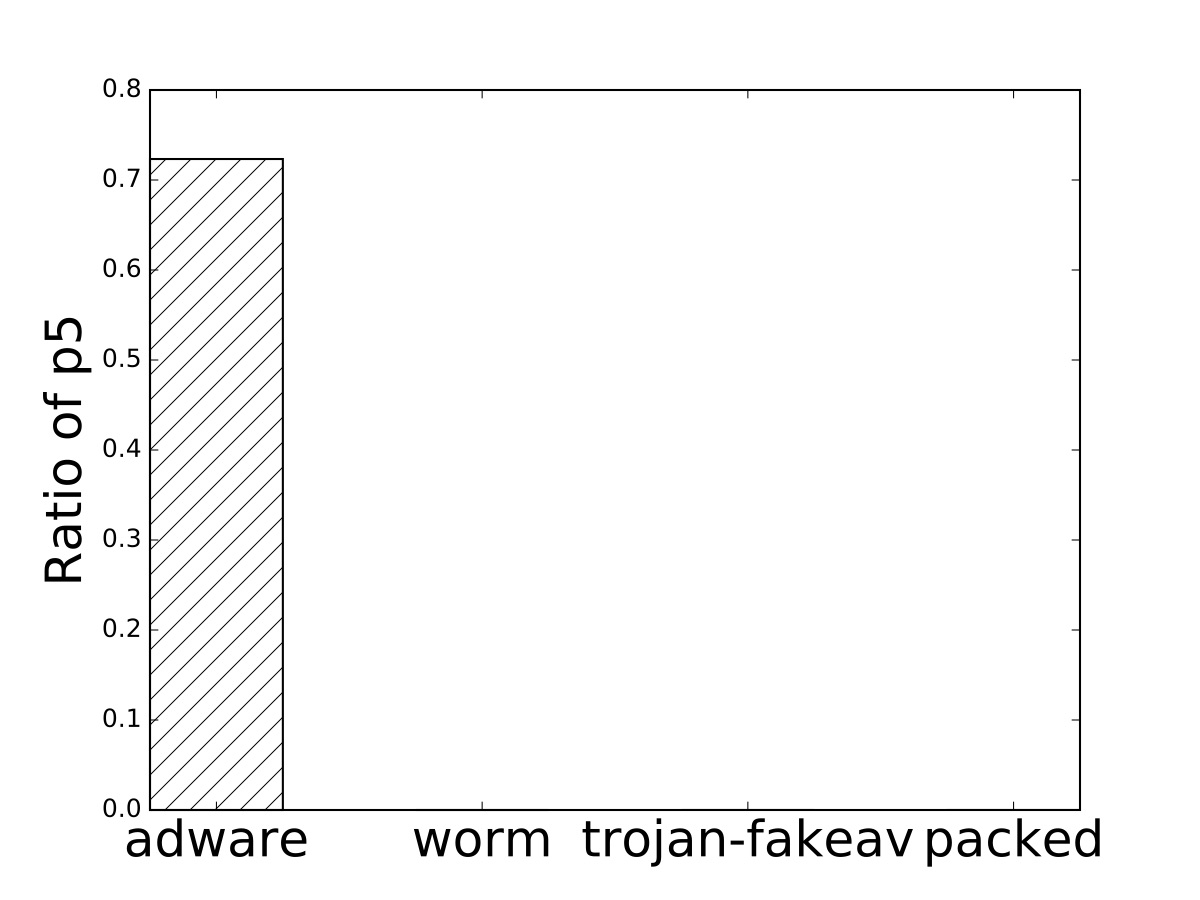}
        \caption{Distribution of FAP p5}
        \label{fig:adware}
    \end{subfigure}
    \caption{Significance of FAPs in Coarse-grained Evaluation}
    \label{fig: coarse-grained distribution}
\end{figure}

\item [Adware] Adware \cite{wiki:Adware} is a form of malware that downloads and displays unwanted ads when a user is online, redirects search requests to certain advertising websites according to the collected user's information without the user's awareness. Generally, an adware will send some HTTP requests according to user's browsing history. Then store the requested files to \textbf{Temp} \cite{web:Temp_folder} directory\footnote{\textbf{Temp} is a default environment variable in Windows system, which points to a path of folder used to store temporary files. Files in Temp folder are absolutely safe to remove.}  for displaying and drop them in the end \cite{analysis:EoRezo_Adware} \cite{analysis:InstallRex}. We can find that the typical subsequence the dominant FAP for adware (see Fig. 3(a) and Tab. 5) has is \textbf{GetTempFileName}, \textbf{SetFileAttributes}, \textbf{DeleteFile}. \textbf{GetTempFileName} create a name for a temporary file and \textbf{SetFileAttributes} is followed to set the related attributes of the file. At last, \textbf{DeleteFile} is called to drop some files. This typical subsequence can be found in around 70\% of the samples of adware and almost none in samples from other families. So the it is highly correlated with behaviours of adware.

\item [Packed \& Trojan-fakeav] Packed malware is a kind of malware that hides the real code of a program through one or more layers of compression or encryption. The top 3 FAPs of \textbf{packed} and \textbf{trojan-fakeav} are similar and do not include $special$ FAPs like \textbf{worm} and \textbf{adware} (see Fig. 3). Packing or more general obfuscation is usually used as an evasive technique to avoid detection and analysis, which is not exclusive to some specific family of malwares. Because a malware packed does not imply specifically deterministic behaviors, so the API call sequence can be quite similar to that of other samples. The reason of the decrease of classification accuracy on $F_3$ and \textbf{packed} arguably is the intersections of the behaviors of the samples. The interweaved samples from different families are not easy to classify, which necessitate the more interpretable information. %\textcolor{red}{*To be adjusted*}

\end{description}

\subsection{Fine-grained Evaluations and Results}
A dataset with 8 families (see Tab. 6) is selected for fine-grained evaluation. We first evaluate the quality of learned representations by visualising with tsne. The quality of the features is quite self-explainatory, samples from different families form different clusters (see Fig. 5). From the results, we can see both the classification and the FAP generation performance are more competitive than that in coarse-grained evaluation (see Tab. 7).

Compared to other families, feature vectors of some samples from \textbf{trojan-fakeav.win32.smartfortress} and \textbf{packed.win32.krap} are quite scattered and interweaved (see Fig. 5). This justifies the essentiality of more insightful and interpretable FAP information, because these samples can be very similar and  are hard to tell apart from the API call sequences.  The fine-grained evaluation also narrow the scope of FAP p6 and p5 to \textbf{net-worm.win32.allaple}  in \textbf{worm} and \textbf{adware.win32.megasearch} in \textbf{adware}, respectively (see Fig. 6).

\begin{figure}[!ht]
    \centering
    %\begin{minipage}{\textwidth}
    %\includegraphics[width=0.45\textwidth]{images/tsne.eps}
    \includegraphics[width=0.75\linewidth]{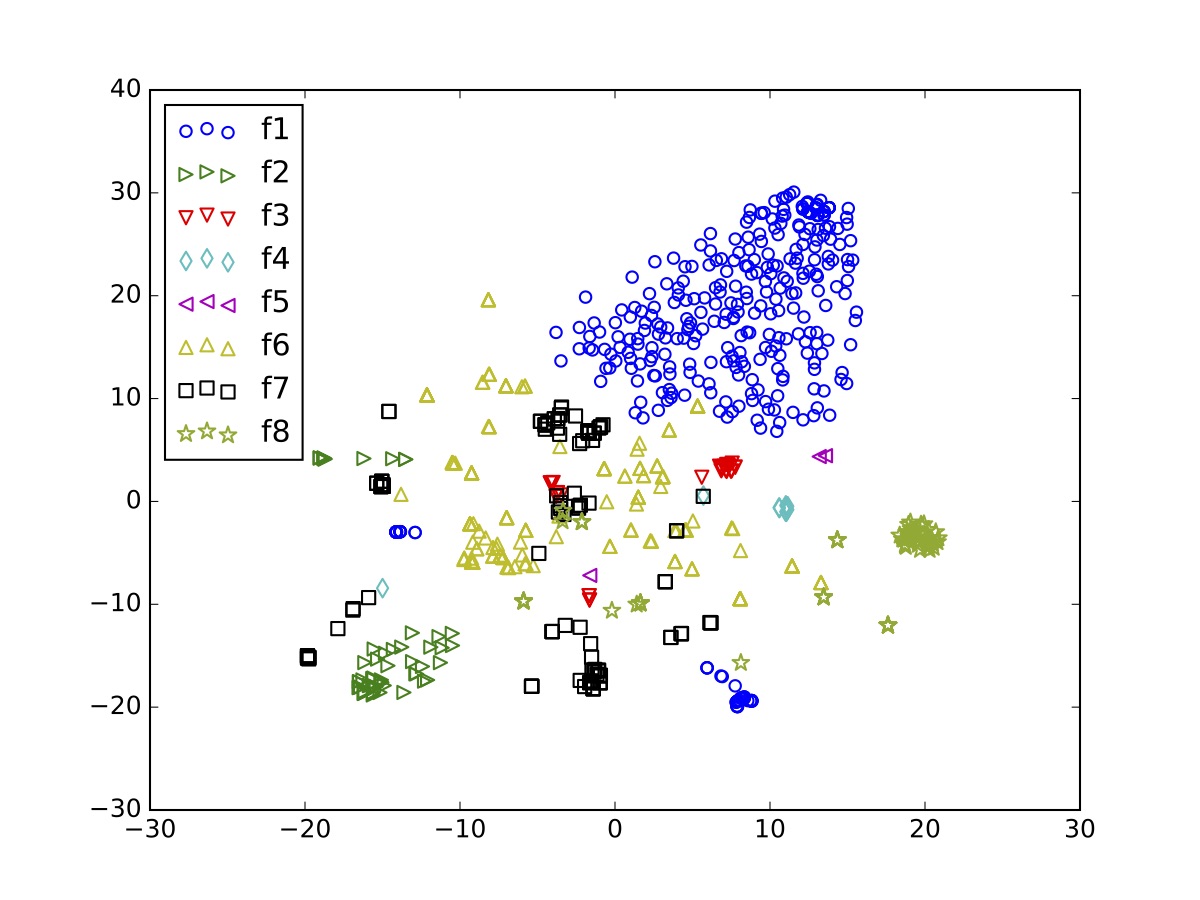}
    %{\footnotesize 30\% of the samples are visualized. \par} 
    %\end{minipage}
    \caption{Visualization of  Malware Features in Fine-Grained Evaluation}
    \label{fig: fine-grained tsne}
\end{figure}

\begin{table}[!ht]
\centering
\resizebox{0.4\textwidth}{!}{\begin{minipage}{0.5\textwidth}
\begin{threeparttable}
\begin{tabular}{ l | c | c | c | c}
  \hline			
  Family & ID & Total & Ratio & Mean\tnote{*}\\
  \hline
  adware.win32.megasearch & f1 & 1644 & 33\% &155\\
  adware.win32.downloadware & f2 & 399 & 8\% & 400\\
  adware.win32.screensaver & f3 & 194 & 4\% & 424\\
  worm.win32.wbna & f4 & 136 &3\% & 470\\
  net-worm.win32.allaple & f5 & 119 & 2\% & 290\\
  trojan-fakeav.win32.smartfortress & f6 & 1205 & 24\% & 217\\
  packed.win32.krap & f7 & 796 & 16\% & 193\\
  downloader.win32.lmn & f8 & 439 & 9\% & 297\\
  \hline
  Total & & 4932 & 100\% & 231\\
  \hline
\end{tabular}
\begin{tablenotes}
  	\item{*} Mean of length of the samples.
\end{tablenotes}
\end{threeparttable}
\end{minipage}}
	\caption{Dataset Statistics and Family Mapping List in Fine-grained Evaluation}
\end{table}

\begin{table}
\centering
\resizebox{0.45\textwidth}{!}{\begin{minipage}{0.5\textwidth}
\begin{tabular}{ l | c | c| c |c }
  \hline			
  \multirow{2}{3cm}{Model} & \multicolumn{2}{c|}{Classification} & \multicolumn{2}{c}{FAP generation} \\
  \cline{2-5}
  	&Train & Test & Train & Test\\ 
  \hline
  seq2seq summarization & 99.9\% & 99.2\% & 99.8\% & 99.3\% \\
  \hline
  AE+Decoder & 99.8\% & 98.9\% & 99.6\% & \textbf{99.2\%} \\
  AE(full)+Decoder & 99.7\% & 99.0\% & 99.7\% & \textbf{ 99.2\%} \\
  \hline
  bAE+ bDecoder & 99.8\% & \textbf{99.1\%} & 99.6\% & 98.9\% \\
  bAE(full)+ bDecoder & 99.8\% & 99.0\% & 99.7\% & 99.0\% \\
  \hline
\end{tabular}
\end{minipage}}
\caption{Fine-grained Evaluation Performance}
\end{table}

\begin{figure}[!ht]
    \centering
    \begin{subfigure}[b]{0.23\textwidth}
        \includegraphics[width=\textwidth]{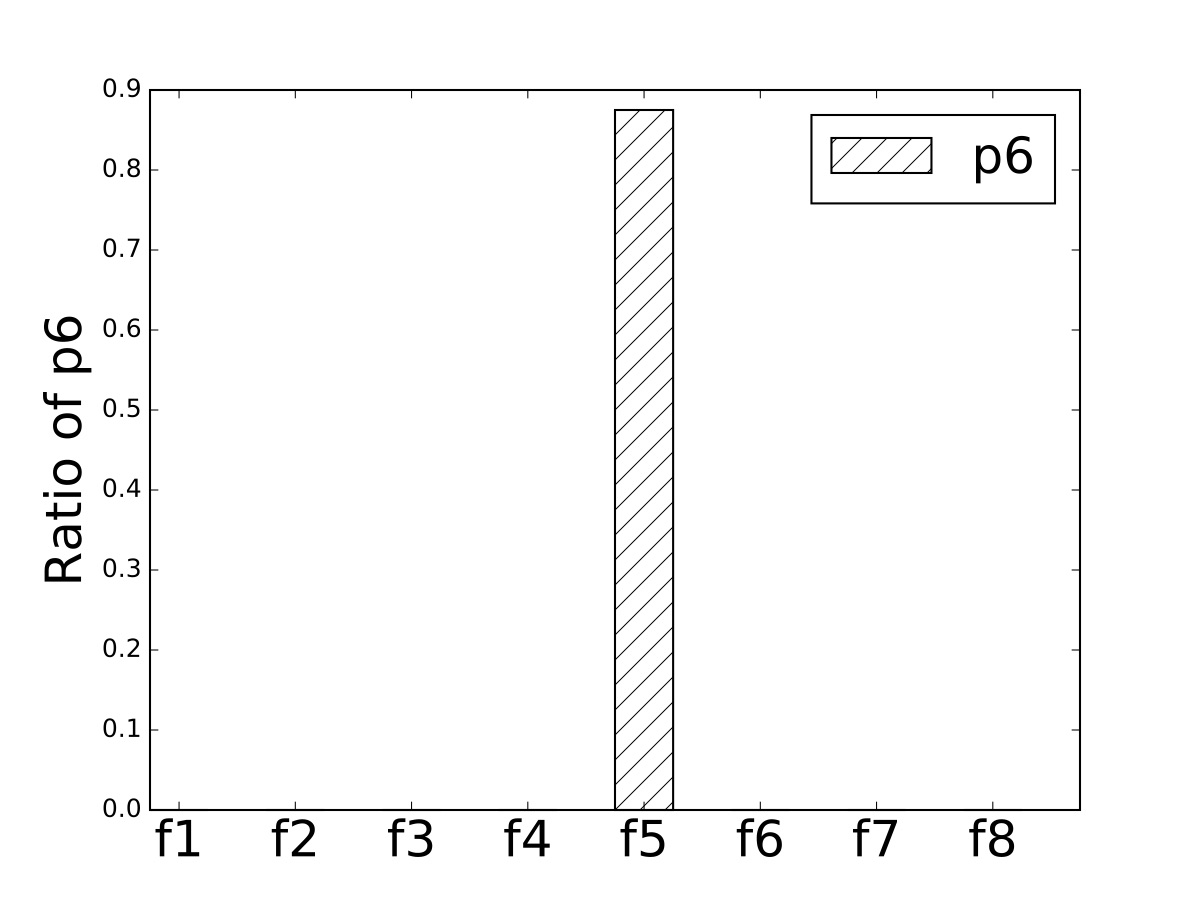}
        \caption{Distribution of FAP p6}
        \label{fig:worm}
    \end{subfigure}
    \begin{subfigure}[b]{0.23\textwidth}
        \includegraphics[width=\textwidth]{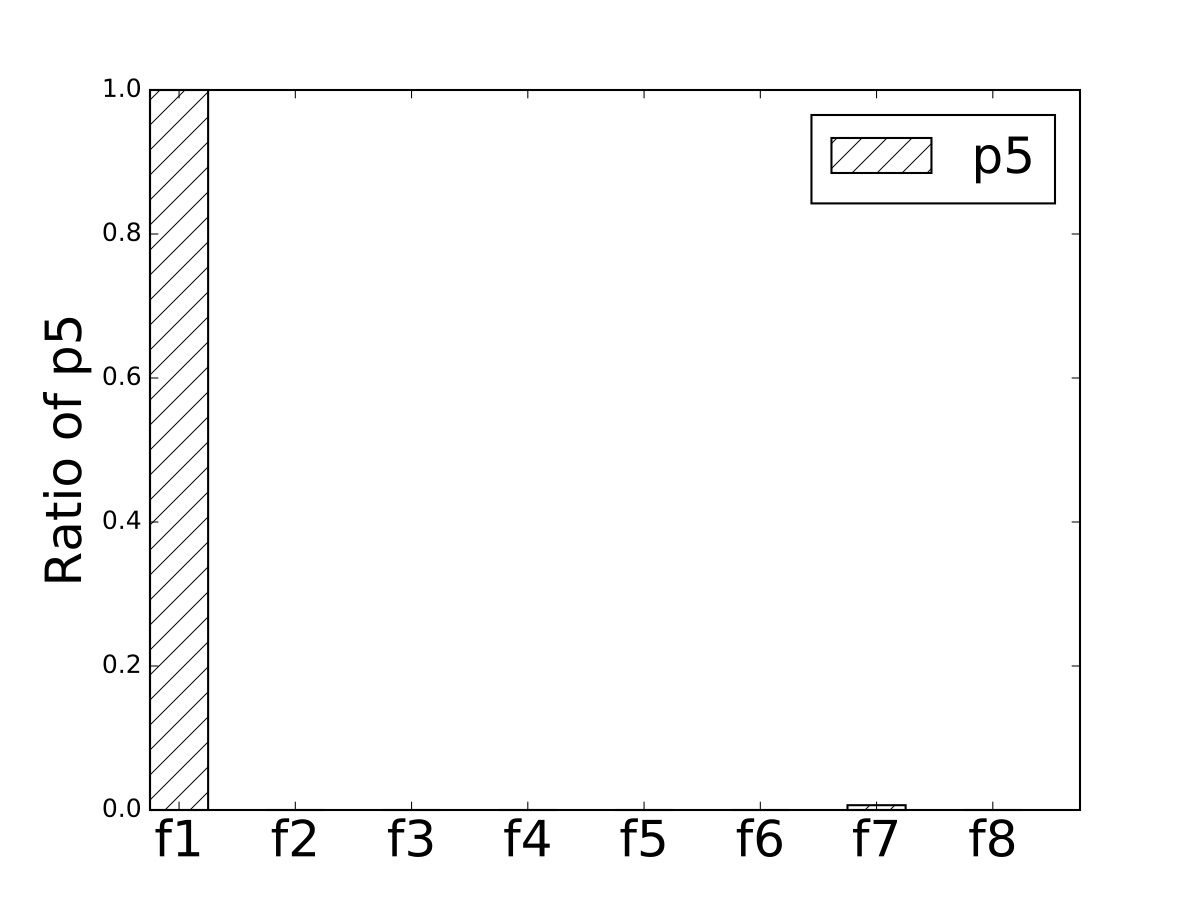}
        \caption{Distribution of FAP p5}
        \label{fig:adware}
    \end{subfigure}
    \caption{Significance of FAPs in Fine-grained Evaluation}
    \label{fig: fine-grained distribution}
\end{figure}

\section{Discussion}
\subsubsection{Scalability}
%Our experiment results show that RNN-AE is able to learn good behavioral aggregation, which can be used for classification and 
%Competitive performance of RNN-AE on unsupervised representation learning makes our model scalable to huge amount of unlabelled malware samples. 

It is not enough that ML-based malware model just give us a class label without any explainatory information. On the one hand, given a malware, a classfier can be wrong, and can be much confident or less confident even if it is right. On the other hand, malwares from the same family do not necessarily perform the same way to our system. It is natural that we want to know what a malware will do to a system except for the family it belongs to.

Experiment results show that our multi-task malware learning model is able to give FAP as well as class label of a malware.  Malware representations learned by RNN-autoencoder from API call sequences are robust enough to trained multiple decoders with quite different objectives and output more informative results.  

Rather than performing one single task at a time, our model first learn representations of malwares in an unsupervised way\iffalse so that it can leverage huge unlabelled data \fi, and then multiple decoders can be trained very efficiently than the training of one single seq2seq task.  

\subsubsection{Limitations and Future works}
So far, we only use API call sequences for representation learning. Different from classification problems of images and texts, diverse source data can be used for classification problems of malware. Apart from API call sequences, so much additional information, like arguments of API call, structure of the executables, can be leveraged for malware detection or classification. How can we merge these kinds of information into our model to improve the classification accuracy and provide more interpretability is to be done in the future.

Another problem is lack of supervision data, because the decoders are trained in a supervised way. In our evaluation on FAP generation, we adopt a very simple way to craft FAPs from API call sequences and prove its effectiveness. Explorations of decoders with new functions and ways to generate corresponding supervision data are very important to build a robust and informative malware learning model. 

\section{Conclusion}
There are two problems of previous works on malware classification: (\rom{1}) Malwares evolve everyday and new unknown families keep emerging. Classifiers built to output known family labels alone are not enough; (\rom{2}) Labels themselves are not very interpretable for samples from the same family may perform quite differently even if the label is right. It is more robust to give a brief description to the behaviors of a malware as well as the class label. We build a multi-task malware learning model based on the proven powerful multi-task seq2seq model for classification and FAP generation. An FAP tells what a malware do to a file system, and sometimes it points directly to the family the malware belongs to. Our tentative results show that not only can seq2seq model be used for malware classification based on API call sequences, but can be used for generating more insightful information from the representations learned by RNN-AE. At the same time, unsupervised representation learning enable the model to automatically leverage huge amount of unlabelled data without further feature engineering. 

\bibliographystyle{aaai}
\bibliography{ref}

\end{document}